\documentclass[doublecol,graphicx]{epl2} 
\usepackage{graphicx}
\usepackage{amsmath}
\usepackage{amsfonts}
\usepackage{amssymb}
\usepackage{amsbsy}

\def\|{{\sss\parallel}}

\def\sss{\scriptscriptstyle}
\title{``Gibbsian" Approach to Statistical Mechanics yielding Power Law Distributions}
\shorttitle{Gibbsian power laws} 

\author{R. A. Treumann\inst{1,2} \thanks{Visiting the International Space Science Institute, Bern, Switzerland} \and W. Baumjohann$^3$}
\shortauthor{R. A. Treumann, \and W. Baumjohann}

\institute{ 
  \inst{1} Department of Geophysics and Environmental Sciences, Munich University, Munich, Germany\\                   
  \inst{2} International Space Science Institute, Bern, Switzerland\\
  \inst{3} Space Research Institute, Austrian Academy of Sciences, Graz, Austria
  }
 

\abstract{Gibbsian statistical mechanics is extended into the domain of non-negligible {though non-specified} correlations in phase space while respecting the fundamental laws of thermodynamics. The appropriate Gibbsian probability distribution is derived and the physical temperature identified. Consistent expressions for the canonical partition function are given. In a first application, the corresponding Boltzmann, Fermi and Bose-Einstein distributions are obtained. It is shown that the latter lose their typical quantum properties, i.e. the degenerate Fermi state and Bose-Einstein condensation. These distributions apply only to states at finite temperature with correlations. As a by-product these results \emph{exclude any negative absolute temperatures} also in the Boltzmann limit.}

\begin{document}

\maketitle
\section{Introduction}
Recently \cite{treumann2008} Gibbsian statistical mechanics was extended to include systems containing some class of internal correlations between particle states described by a piecewise constant positive real number `ordering parameter' $\kappa \in\textsf{R}$, considered a free parameter. Original motivation for this theory was taken from the continuous observation of power law distributions in collisionless space plasma, exhibiting high energy tails on the distribution of particles with respect to velocity $v$ and energy $\epsilon=m\gamma c^2$ (or nonrelativistically $\epsilon=mv^2/2$). Such $\kappa$-distributions have been used in fitting observed particle energy fluxes since their introduction by Vasyliunas \cite{vasyliunas1968} following a suggestion of their mathematical form by S. Olbert (cf. the acknowledgement in \cite{vasyliunas1968}). Distributions of this kind are in fact Cauchy-Lorentzians \cite{treumann1999a} generalized to real powers $\kappa>0$, which for $\kappa\to\infty$ smoothly approach Maxwell-Gaussians. {They are closely related to the non-extensive Statistical Mechanics proposed by Tsallis} \cite{tsallis1988} based on earlier attempts \cite{jack1905,hahn1949,reny1956,dar1970,reny1970} of $q$-generalizations constructing functional dependences for the entropy $S_q[p]$ on probability $p$ other than the celebrated Boltzmann-Shannon logarithmic postulate. The free index $q$ appearing in non-extensive Statistical Mechanics is {formally related to $\kappa$ via the simple transformation $q=1+1/\kappa$\cite{treumann1999a}} widely used in the literature on  $\kappa$ distributions, with limit $\kappa\to\infty$ corresponding to $q\to1$. {In the absence of binary collisions, internal correlations are provided by interaction between particles and  turbulence yielding functional dependences $\kappa[W_w]$ on the spectral energy density $W_w$ of the turbulence\footnote{The implicit dependence $\kappa[W_w]$ on the self-consistent wave turbulence level first proposed in \cite{treumann1999b} also implies $q[W_w]$ in non-extensive thermo-statistics thus giving physical meaning to both parameters as quantification of the internal correlations in phase space.}  found for electron distributions embedded into a photon bath \cite{hasegawa1985}\footnote{{This reference just finds the exponent of the high-energy power law tail on a Maxwell distribution as function of wave power, resulting from solution of a diffusion equation, different from the intention of this paper. Particle number and energy conservation require such power laws to be exponentially truncated at some limit energy. }} and, more specifically, in electron interaction with thermal plasma wave levels accounting for spontaneous emission, scattering and absorption \cite{yoon2012}, the latter providing a firm basis for generalized Cauchy-Lorentzians in this particular case.}

The Gibbsian approach given in \cite{treumann2008} was based on the known mathematical generalization of the exponential  (see, e.g., \cite{livadiotis2013}). Application to  particle phase space resulted in the canonical particle probability distribution $w_{i\kappa}(\epsilon_i)$ of finding a particle in energy state $\epsilon_i$. {The distribution still depended on} the total entropy $S(E)$ (taken in units of the Boltzmann constant $k_B$) of the system of particles with total energy $E$ which was unfortunate as it prevented giving the temperature its ordinary thermodynamic interpretation. In the present note {this problem is resolved}. The theory correctly reproduces the thermodynamic relations{, becoming} a viable statistical mechanics. 

\section{Gibbsian approach}
In Gibbsian theory{\footnote{{Speaking of ``Gibbsian approach" follows the Gibbsian intention of basing the  state of a system in thermal equilibrium on the evolution of the occupied phase space, the latter being determined by the entropy $S(\epsilon)$ as function of particle energy $\epsilon$. Gibbsian theory implies a strict thermal equilibrium. Hence we also assume the existence of a thermal equilibrium as a formal approach. Its limitations will be discussed below.}}}  the probability $w_i(\epsilon_i)\propto\exp(-\epsilon_i/T)$ of finding a particle in energy state $\epsilon_i$ at constant temperature $T$ (in energy units) is obtained from a consideration of the evolution of a finite phase space volume $\Gamma(\epsilon)$ embedded into an extended heat bath. The infinitesimal change in $\Gamma$ is given by the entropy $S(\epsilon)$ as $\mathrm{d}\Gamma/\mathrm{d}\epsilon=\exp[S(\epsilon)]/\Delta\epsilon$. With the generalised Lorentzian substitution for the exponential{\footnote{{Generalized Lorentzians $\left(1+x/\kappa\right)^{-\kappa}$  (with arbitrary expression $x$ independent on $\kappa$) are not the only functions having the correct exponential limit. Depending on what function is chosen, different generalizations of Gibbs' approach are obtained. Those reproducing the thermodynamic relations yield a physically acceptable Statistical Mechanics.}}} 
this change becomes
\begin{equation}
\frac{\mathrm{d}\Gamma(\epsilon)}{\mathrm{d}\epsilon}=\frac{1}{\Delta\epsilon}\left\{1+\frac{1}{\kappa}\left[S(E)-S(\epsilon)\right]\right\}^{-\kappa-r}
\end{equation}
It is obvious that this expression for $\kappa\to\infty$ reproduces the Gibbsian variation of the phase space volume. The negative sign in the brackets is self-explanatory. The constant total entropy $S(E)$ has been included which for $\kappa\to\infty$ gives just a constant factor in d$\Gamma$. We also made use of the freedom of adding a number $r\in\textsf{R}$ to the exponent, as it does not change anything when taking the large $\kappa$ limit. {That finite $\kappa<\infty$ imply correlations in phase space is clear from destruction of the exponential dependence of the phase space volume element on entropy $S$. It breaks the assumption of additivity of entropy that underlies ordinary Gibbsian Statistical Mechanics.\footnote{{This is seen when (for simplicity even with constant $\kappa$) considering the phase space volume element d$\Gamma_3$=d$\Gamma_1$d$\Gamma_2$, with $S_3=S_1+S_2$, trivially yielding d$\Gamma_3=\{1+[S_3(E)-S_3(\epsilon)]/\kappa +[S_1(E)-S_1(\epsilon)][S_2(E)-S_2(\epsilon)]/\kappa^2\}^{-\kappa-r}$. In contrast to Gibbs' case, the correlation integral over d$\Gamma_3$ thus differs from the product of integrals over the two phase space elements d$\Gamma_1$=d$q_1$d$p_1$ and d$\Gamma_2$=d$q_2$d$p_2$. It contains an additional (irreducible quadratic) term indicating that the two phase space elements in the $\kappa$-generalized Gibbsian model are not independent; they are correlated. Since reduction of $\kappa$ from $\infty$ to $\kappa<\infty$ is the only change made, $\kappa$ bears responsibility for the (unspecified) correlations in phase space and deviations from } {stochasticity. This violates the ergodic hypothesis. \emph{Finite $\kappa$ implies that not all regions of phase space are visited with same probability}. Regions of high energy are visited excessively often, as indicated by the high energy tail in the resulting probability $w_\mathit{ik}(\epsilon_i)$ in Eq. (\ref{gibbsdist}). }}}

When expanding the entropy around $E$ with respect to energy $\epsilon_i$ in state $i$, we obtain
\begin{equation}\label{entexp}
S(\epsilon)\sim S(E)+\frac{\mathrm{d}S(\epsilon)}{\mathrm{d}\epsilon}\bigg|_{\epsilon=0}(\epsilon-E)=S(E)+\frac{\epsilon-E}{T} 
\end{equation}
with $\partial S(E)/\partial E=1/T$. The probability of finding the particle in state $i$ then follows from the integral 
\begin{equation}
w_i\propto \int\left[\frac{\mathrm{d}\Gamma(\epsilon)}{\mathrm{d}\epsilon}\right]\delta(\epsilon_i+\epsilon-E)\ \mathrm{d}\epsilon
\end{equation}
Inserting for the variation of the phase space volume and integrating yields
\begin{equation}\label{gibbsdist}
w_{i\kappa}(\epsilon_i,\mathbf{x})=A\Big\{1+{\big[\epsilon_i+\Phi(\mathbf{x})\big]}/{\kappa\ T}\Big\}^{-\kappa-r}
\end{equation}
 an expression similar to that obtained in \cite{treumann2008} for the canonical probability distribution $w_{i\kappa}(\epsilon)$ in state $i$ as function of temperature and entropy but independent on total entropy, generalised to non-ideal gases by including a potential energy $\Phi(\mathbf{x})$.  $A$ is a constant of normalization of the probability when integrating over phase space d$\Gamma$. Eq. (\ref{gibbsdist}) allows for a formulation of Gibbsian statistical mechanics consistent with fundamental thermodynamics.

To determine the value of $r$, we switch temporarily from probability to phase space distribution $w_{i\kappa}(\epsilon_i)\to f_\kappa(\mathbf{p})$, with $\mathbf{p}$ particle momentum and $\epsilon(\mathbf{p})=p^2/2m$ and restrict to ideal gases. Note that normalization $(2\pi\hbar)^{-3}\int f_{\kappa}(\mathbf{p}) \mathrm{d}\Gamma$ integrated over all physical phase space determines $A_{f_\kappa}$ as function of particle number $N$ and volume $V$ to
\begin{equation}
A_{f_\kappa}=\frac{N\lambda_T^3}{V}\frac{\kappa^{-\frac{3}{2}}\Gamma\big(\kappa+r\big)}{\Gamma\big(\frac{3}{2}\big)\Gamma\big(\kappa+r-\frac{3}{2}\big)}
\end{equation}
where $\lambda_T=\sqrt{2\pi\hbar^2/ mT}$ is the thermal wavelength. The average energy is obtained from the kinetic energy moment of the normalized distribution function.This produces an additional factor $(\kappa T)^{\frac{5}{2}}\Gamma(\frac{5}{2})\Gamma(\kappa+r-\frac{5}{2})/\Gamma(\kappa+r)$. Combination yields for the ideal gas kinetic energy density
\begin{equation}\label{eq-eos}
\frac{\langle E\rangle}{V} = \frac{3}{2}\frac{\kappa}{\kappa+r-\frac{5}{2}}\frac{NT}{V}
\end{equation}
For three degrees of freedom of the particles, it is required that $\langle E\rangle=\frac{3}{2}\,NT$ which immediately yields that 
\begin{equation}
r= {\textstyle\frac{5}{2}} 
\end{equation}
(Correspondingly, for $k$ degrees of freedom one had $r=1+\frac{1}{2}k$.) Hence the phase space distribution becomes
\begin{equation}\label{eq-gibbs}
f_{\kappa}(\mathbf{p})=A_{f_\kappa}\left(1+\frac{\epsilon_i}{\kappa\ T}\right)^{-\kappa-\frac{5}{2}}
\end{equation}
In addition it proves that $T$ is the correct physical temperature of the system, which is a parameter characterizing its thermal state.{\footnote{{In particular, the $\kappa$ gas embedded into a heat bath will assume the temperature of the heat bath. Also two gases in contact at a boundary will after a while assume the same temperature, independent on whether both are $\kappa$ gases of equal or different $\kappa$ or one of them being a Boltzmann gas.}}} This had been shown earlier in the Generalized Lorentzian thermodynamics \cite{treumann1999b} though the correct value of $r$  had not been obtained yet.\footnote{The correct $r$ value was inferred in \cite{livadiotis2009} from inspection of the classical particle distribution obtained in non-extensive $q$-statistics for the case $\mu=0$ (which holds for systems of fixed particle number) by arguing about the role of total energy $E$. It was also obtained from a direct kinetic theoretical calculation \cite{yoon2012} of the particular case of the \emph{time-asymptotic limit} of the electron state far from thermal equilibrium in interaction between electrons and the thermal Langmuir wave background. {These latter calculations show that any interaction leading to power law distributions is basically non-stationary thereby confirming numerical simulations. Statistical mechanics based on power laws is a temporarily quasi-stationary theory in slow evolution holding for times substantially shorter than binary collision times.} For a tabulated compilation of the various properties of the classical non-extensive $\kappa$-particle distribution the reader is referred to\cite{livadiotis2013}.} This is a very important assertion as it assigns the temperature $T$ its well-known thermodynamical meaning.

The value of $r$ indeed passes through from the Gibbs distribution to the particle distribution function thus a posteriori physically justifying the heuristic approach in \cite{livadiotis2009}. That this follows directly from fundamental Gibbsian theory is an important fact. It demonstrates that the property of super-extensivity is not simply a matter of another version of distribution function or entropy. Rather it is the consequence of the phase space underlying statistical mechanics. Under the conditions leading to the generalized Gibbsian statistical mechanics the particles are correlated in phase space, implying that the {particle-}phase space elements are not independent. {This might, for instance, be the case under conditions of turbulence when, in the absence of binary particle collisions, interactions between particle population and turbulent wave spectrum affect the particle dynamics, rendering the parameter $\kappa[W_w]$ a functional of the quasi-stationary turbulent wave intensity.}

Including a variable particle number $N$, the entropy $S(E,N)$ becomes in addition a function of $N$. Expanding $S$ with respect to energy and particle number, defining the derivative of the entropy at constant energy ($\partial S/\partial N)_{N=0,E}=-\mu/T$, the extended Gibbs distribution is given by
\begin{equation}
w_{i\kappa}(\epsilon_{iN})=A\left(1+\frac{\epsilon_{iN}-\mu N}{\kappa\ T}\right)^{-\kappa-r}
\end{equation}
The index $N$ identifies the $N$th subsystem of fixed particle number $N$. (If the subsystems contain just one particle, it is the particle number.) This is the general form of the $\kappa$-Gibbs probability distribution with physical temperature $T$ in state $(iN)$.

\section{Generalized Gibbsian statistical mechanics}
The form of the Generalized Gibbsian distribution inhibits the use of the Gibbsian definition of entropy $S=-\langle \log w\rangle$ as phase space average $\langle\dots\rangle\equiv \int \dots d\Gamma$ of the logarithm of $w(\epsilon)$. Instead, another form of entropy has to be found enabling construction of a generalized thermodynamics in agreement with the fundamental thermodynamic definitions. Such forms have been proposed in $q$ statistical mechanics \cite{tsallis1988, gell-mann2004} and also for the Generalized Lorentzian thermodynamics \cite{treumann1999b}. We adopt the latter version defining a functional $g[w]$  whose logarithmic average leads to the entropy $S=-\big\langle\!\log g[w]\big\rangle$. The function $g[w]$ is given as
\begin{equation}
g[w]=\exp\left\{-\kappa\left[\bigg(\frac{A}{w_{\kappa}}\bigg)^{(\kappa+r)^{-1}}\!\!\!\!\!\!-1\right]-\log{A}\right\}
\end{equation}
Its particular form is chosen for reconciling with thermodynamics. In addition a normalization factor $A$ has been added. Substituting the generalized Gibbs distribution $w_{i\kappa}(\epsilon_i)$ and $g[w_{i\kappa}(\epsilon_i)]$ into the phase space average yields
\begin{equation}
S = -\log {A} +\frac{\langle E\rangle}{T}
\end{equation}
which by the thermodynamic relation $\langle E\rangle=TS+F$ identifies  $F =T\log A$ as the free energy $F$. Hence the generalized canonical Gibbs distribution may be written in the form
\begin{equation}
w_{i\kappa}=\frac{\exp{(F/T)}}{\left(1+\epsilon_i/\kappa\ T\right)^{\kappa+r}}
\end{equation}
Since $A$ is the normalization of $w_{i\kappa}$ one also has that $\sum_iw_{i\kappa}=1$ and, hence, for the free energy
\begin{equation}
F=-T\log \int \mathrm{d}\Gamma \bigg[1+\frac{\epsilon(\mathbf{p,x})}{\kappa\ T}\bigg]^{-\kappa-r} = -T\log Z_\kappa
\end{equation}
with d$\Gamma=\mathrm{d}^3p\,\mathrm{d}V/(2\pi\hbar)^3$ the phase space volume element. From the last expression we can immediately read the generalized Gibbsian version of the classical canonical partition function
\begin{equation}
Z_\kappa\equiv \int \mathrm{d}\Gamma \bigg[1+\frac{\epsilon(\mathbf{p,x})}{\kappa\ T}\bigg]^{-\kappa-r}
\end{equation}
We just remark that in the quantum case the integral becomes a sum over all quantum states $i$ as
\begin{equation}\label{eq-z}
Z_\kappa\equiv\sum_i\bigg(1+\frac{\epsilon_i}{\kappa\ T}\bigg)^{-\kappa-r} 
\end{equation}
This completes our discussion for a system with fixed particle number since all statistical mechanical information is contained in the partition function $Z_\kappa$.

Allowing for a variable particle number $N$ and again making use of  the chemical potential $\mu=-T(\partial S/\partial N)_{EV}$, the normalization condition becomes
\begin{equation}
T \log A =F-\mu \langle N\rangle \equiv \Omega
\end{equation}
with $\langle N\rangle$ the average particle number, and $\Omega$ the thermodynamic potential. With $F=\mu N +\Omega$ for the $N$th subsystem of particle number $N$ the generalized Gibbs distribution  reads
\begin{equation}
w_{\kappa, N}=\frac{\exp(\Omega_\kappa/T)}{\big[1+(\epsilon_{iN}-\mu N)/\kappa\ T\big]^{\kappa+r}}
\end{equation}
repeating that $N$ is the index of the subsystems of different particle number. Normalization requires in addition summing  over these $N$ subsystems, a procedure not affecting $\Omega$ and thus yielding
\begin{eqnarray}
\Omega_\kappa&=& -T \log\sum_N\sum_i\bigg[1+\big(\epsilon_{iN}-\mu N\big)/\kappa\ T\bigg]^{-\kappa-r} \cr
&=&-T\log Z_\kappa
\end{eqnarray}
The argument of the logarithm is the grand partition function $Z_{\kappa}=\sum_N  Z_{\kappa, N}$ which is the sum over all partition functions of the $N$ separate subsystems of different particle numbers $N$. For classical systems the inner sum over $i$ becomes again the $N$th phase space integral. Otherwise it is the sum over states. All thermodynamic information is contained in it. 

The most interesting case is that of $N$ undistinguishable subsystems (particles) in states $i$. Then the sum over $N$ in the logarithm is understood as an exponentiation yielding for the free energy
\begin{equation}
F=-NT\log\int\frac{\mathrm{d}\Gamma\exp(1)/N(2\pi\hbar)^3}{\Big[1+\big(\epsilon_\mathbf{p}-\mu\big)/\kappa\ T\Big]^{\kappa+r}}
\end{equation}
an expression to be used for the determination of the common chemical potential $\mu$ from its definition $\partial F/\partial N=\mu/T$. This yields, with $r=\frac{5}{2}$, the implicit equation for $\mu<0$
\begin{equation}\label{eqmu}
\frac{\mu}{\kappa\,T}=\log\left[\frac{V\exp(1)}{N\lambda_\kappa^3}\frac{\mathrm{B}\left(3/2,\kappa+1\right)}{\left(1-\mu/\kappa\,T\right)^{(\kappa+1)}}\right]
\end{equation}
The logarithmic dependence on $\mu$ is weak. Hence the chemical potential of the ideal classical gas of undistiguishable particles  is essentially the same as that for the ordinary ideal Boltzmann gas. However, in contrast to the latter, $\mu$ is a substantial part of the distribution function which cannot be absorbed into the normalization. 

\subsection{Boltzmann-$\kappa$  distribution} Following Gibbsian philosophy we consider particles in a given state $i$ and write for the index $N=n_i$, with $n_i$ the occupation number of state $i$, and for the energy $\epsilon_{iN}=n_i\epsilon_i$. Then one has for
\begin{equation}\label{eq-thermpot}
\Omega_{\kappa i}=-T \log \sum_{n_i}\bigg[1+n_i\frac{(\epsilon_i-\mu)}{\kappa T}\bigg]^{-\kappa-5/2}
\end{equation}
The Gibbsian probability distribution for the occupation numbers thus becomes
\begin{equation}
w_{\kappa n_i}=\exp(\Omega_{\kappa i}/T)\big[1+n_i(\epsilon_i-\mu)/\kappa T\big]^{-(\kappa+5/2)}
\end{equation}
The probability for a state to be empty is obtained for $n_i=0$. Thus $w_{\kappa 0}=\exp(\Omega_{\kappa i}/T)$ is identical to the non-$\kappa$ case of zero occupation. At high temperatures the occupation numbers of states are generally small. Hence the average occupation $\langle n_i\rangle$ is obtained for $n_i=O(1)$ and $\Omega_{\kappa i}/T\ll 1$, yielding
\begin{equation}\label{boltz}
\langle n_i\rangle =\sum_{n_i} n_iw_{\kappa n_i}\approx w_{\kappa 1}\approx \big[1+(\epsilon_i-\mu)/\kappa T\big]^{-\kappa-5/2}
\end{equation}
This average occupation number is the $\kappa$-equivalent of the Boltzmann distribution of occupation numbers of states in an ideal gas with no external interaction potential but variable particle number as obtained from rigorous Gibbsian theory, with $\mu$ given in Eq. (\ref{eqmu}) a negative number. Inspection suggests that this distribution of occupations is about flat for $\epsilon_i<|\mu|$. For constant particle number it instead becomes Eq. (\ref{eq-gibbs}), the ordinary (canonical) one-particle $\kappa$ distribution \cite{vasyliunas1968,livadiotis2013}.  {In going from occupation of states to the distribution function, normalization is to the particle density $N/V$. Extracting $1+|\mu|/\kappa T$ from the integral, the contribution of $\mu$ can, like in the Boltzmann case, be absorbed into the normalization constant, with energy in units of $\kappa T + |\mu|$.} 

In addition, any interaction potential $\Phi(\mathbf{x})$ at location $\mathbf{x}$ affecting particle dynamics in non ideal gases must be added to the energy in the distribution function which, in this case, becomes dependent on real space. 

\subsection{Fermi-$\kappa$ distribution of occupation numbers}
From the general expression Eq. (\ref{eq-thermpot}) of the Gibbs thermodynamic potential it is possible to obtain the distribution of occupation numbers under various conditions. The first is the Fermi assumption that any energy states can host at most one particle. Hence, their occupation numbers are restricted to $n_i=0,1$. This gives Fermi's version 
\begin{equation}
\Omega_{\kappa i}^{F}=-T\log\bigg\{1+\bigg[1+\frac{\big(\epsilon_i-\mu\big)}{\kappa T}\bigg]^{-(\kappa+5/2)}\bigg\}
\end{equation}
The average occupation  number of states follows from here as
\begin{equation}\label{fermi}
\langle n_i\rangle_{\kappa}^F=-\frac{\partial\Omega_{\kappa i}^F}{\partial\mu}=\bigg[1+\bigg(1+\frac{\epsilon_i-\mu}{\kappa T}\bigg)^{\kappa+5/2}\bigg]^{-1}
\end{equation}
which has to be normalized by the condition that summation $\sum_i$ over all states $i$ yields the total particle number $N$. 

It is instructive to consider the distribution in the limit $T\to0$. For $\epsilon_i>\mu$ both $\Omega_{\kappa i}^F$ and $\langle n_i\rangle_\kappa^F$ vanish when $T\to0$. On the other hand, any energy level below $\mu$ cannot be occupied when the temperature vanishes for the reason that the distribution must be real.  

Hence, the zero temperature limit is not accessible to the $\kappa$-Fermi distribution, which exists only at finite temperatures. This is in accord with the idea that correlations in phase space imply some complicated dynamics and, hence, finite temperature. From here it follows that the Fermi-$\kappa$ distribution does not define any Fermi energy since no degenerate states exist. At $T>0$ any positive chemical potential would be bound by $\mu<\kappa T$ not providing any new information.

\subsection{Fractal Fermi-$\kappa$ distribution}
Gibbsian theory also allows for a fractal occupation of states in the Fermi case. This can be constructed when assuming that any energy states $i$ may exhibit fractal occupations in the sense that a state $i$ can be accessible not only for the two occupations $n_i=0, 1$ but also to any fraction $s_i/\ell_i$ with $0\leq s_i\leq\ell_i$ and $\ell_i\in\textsf{R}$. For $s_i=0$ the state $i$ is empty, while for $s_i=\ell_i$ the state contains just 1 particle. If particles can manage to share fractionate their energy  and share it between different states, then any intermediate state $0<s_i<\ell_i$ will be occupied fractionally. For such a state the thermodynamic potential becomes
\begin{equation}
\Omega_{\kappa i}= -T\log\sum_{s_i=0}^{\ell_i}\big[1+s_i(\epsilon_i-\mu)/\kappa\ell_iT\big]^{-\kappa-5/2}
\end{equation}
The sum in this expression cannot be easily performed. One can, however, find a version of the average occupation number by taking the derivative with respect to $\mu$ in this expression. This yields
\begin{equation}
\langle n_i\rangle_{\kappa, \ell_i}= \frac{1}{\ell_i}\frac{\left(1+\frac{5}{2\kappa}\right)\sum\limits_{s_i=1}^{\ell_i}s_i\Big[1+s_i\big(\epsilon_i-\mu\big)/\kappa\ell_iT\Big]^{-\kappa-7/2}}{1+\sum\limits_{s_i=1}^{\ell_i}\Big[1+s_i\big(\epsilon_i-\mu\big)/\kappa\ell_iT\Big]^{-\kappa-5/2}}
\end{equation}
This distribution does not resemble the Fermi distribution. Rather it resembles the Bose distribution derived below. This is the effect of summation over many more than just two terms in the nominator and denominator. We note that a better suited form abandoning the untreatable sums can be constructed by applying the method used below in simplifying the Bose-Einstein distribution.

\subsection{Bose-Einstein-$\kappa$ distribution of occupation numbers}
When the occupation number of states is arbitrary, summation over all states $n_i$ from $n_i=0$ to $n_i=\infty$ is in place in the Gibbs thermodynamic potential. One immediately realizes that  the chemical potential $\mu<0$ must be negative. 

The Bose-Einstein-$\kappa$ distribution is obtained by taking the derivative with respect to $\mu$
\begin{equation}
\langle n_i\rangle_{\kappa}^{BE}= \left(1+\frac{5}{2\kappa}\right)\frac{\sum\limits_{n_i=1}^{\infty}n_i\Big[1+n_i\big(\epsilon_i-\mu\big)/\kappa T\Big]^{-\kappa-7/2}}{1+\sum\limits_{n_i=1}^{\infty}\Big[1+n_i\big(\epsilon_i-\mu\big)/\kappa T\Big]^{-\kappa-5/2}}
\end{equation}
In contrast to the Fermi-$\kappa$ distribution this expression cannot anymore be brought into closed form. 

The low temperature behavior of this distribution is again completely determined by the chemical potential $\mu$. It is readily shown that in the low temperature limit $T\to0$ the distribution vanishes completely for all states $\epsilon_i\neq 0$ and for all $\mu<0$. Moreover, there is no Bose-Einstein condensation on the lowest energy level $\epsilon_0=0$. This can easily be demonstrated by taking the limit $\mu\to0$ and $T\to0$. Hence, the above distribution applies to systems of finite temperature only.

In order to simplify the distribution one needs to refer to approximations. One of those simplifications can be obtained by adopting a procedure frequently used in quantum physics, i.e. temporarily replacing all expressions of the kind $(1+x)\to \exp x$. Then all sums containing exponentials can be performed using the summation rules for geometric progressions. Resolving all exponentials in the final result by the original expressions, one obtains the wanted approximation for the Bose-Einstein-$\kappa$ distribution
\begin{eqnarray}\label{be}
&\langle n_i\rangle_{\kappa}^{BE}&\approx{\bigg\{1-\Big[1+(\epsilon_i-\mu)/\kappa T\Big]^{-\kappa-{\frac{5}{2}}}\bigg\}}\times \cr
&\times\!\!\!\!\!\!\!&\!\!\!\!\!\!\left[\frac{(1+{\textstyle\frac{5}{2}}\kappa\Big)\Big[1+(\epsilon_i-\mu)/\kappa T\Big]^{-\kappa-\frac{7}{2}}}{\bigg\{1-\Big[1+(\epsilon_i-\mu)/\kappa T\Big]^{-\kappa-\frac{7}{2}}\bigg\}^{\!2}}\right]
\end{eqnarray}
which may be more appropriate for applicational purposes.

\section{Conclusions}
In the present note we revisited the generalized Gibbsian formulation of a ``stationary statistical mechanics far from thermal equilibrium'' with ordering parameter $\kappa$ formally accounting for the presence of internal phase space correlations. {The nature of correlations remains unspecified. It is  assumed that they mediate the interaction between the particles and dominate over the binary collisions (stochastic processes) on which conventional Gibbsian theory is based. The theory is thus appropriate for collisionless systems like dilute high temperature plasmas.}

{Correlations in particle phase space can be produced by wave-particle interactions. They imply that the interaction between particles proceeds via excitation of and scattering in turbulent field fluctuations, processes well-known in collisionless plasmas. Such interactions violate the ergodic hypothesis, i.e. the states in phase space are not visited with same probability. The canonical probability obtained indicates that higher energy states are subject to a higher non-stochastic visitation rate. In the present theory it is assumed that wave-particle interactions are responsible for values of $\kappa<\infty$. This means that $\kappa[W_w]$ itself is a functional of the turbulent wave power. This functional dependence has been left unspecified, but direct calculations \cite{hasegawa1985,yoon2012} for two particular cases have verified this assumption providing fairly involved expressions for $\kappa[W_w]$.}

{In contrast to the ergodic Gibbsian statistical mechanics such a theory is not strictly stationary. Turbulent interactions  involve dissipation on the microscopic scale. Stationary states require continuous supply of energy and some kind of cooling (losses due to heat flux) to adjust to stationary temperatures. The theory holds for open systems where energy supply and losses are well balanced. In their absence, the assumption of stationarity is valid only for times less than binary collision and heating times. Within these limits the system is considered quasi-stationary and in quasi-thermal equilibrium. This seems to be the case in cosmic ray physics and, as for another example, also in the physics of the heliosphere \cite{christon1988,christon1991,fisk2006} where the energetic particle populations obey marginally stable $\kappa$ distributions. A more general Gibbsian approach should include wave turbulence and boundaries to allow for supply and loss of energy. One way to achieve this goal is enlarging the phase space to include wave momenta.}

The expressions {in the present paper} were adjusted to the requirements of the fundamental thermodynamic relations. {Any two systems in contact with each other will necessarily assume the same temperature, independent on whether they are $\kappa$ gases or not or have different $\kappa$s.} We constructed the quantum distributions.
The Fermi-$\kappa$ distribution observes no degenerate state at low temperatures, which is expected in the absence of correlations at $T=0$. The Bose-Einstein-$\kappa$ distribution contains no zero-temperature state thus not allowing for condensation at zero temperature. The Boltzmann-$\kappa$ distribution also becomes invalid at vanishing temperature. Non-extensive Gibbsian statistical mechanics excludes vanishing absolute temperatures for, at zero temperature, no correlations affecting $\kappa[W_w]$ are produced. 

In particle systems obeying $\kappa$ states, any negative absolute temperatures $T<0$ are impossible. They would require cooling across the non-existing state $T=0$. Since $\kappa\to\infty$ reproduces classical thermodynamics for all $T$, this conclusion provides another proof following from Gibbsian theory for the \emph{nonexistence of negative} absolute temperatures, which supports and adds to a recent proof \cite{dunkel2014}.

{\small \acknowledgements
This research was part of a Visiting Scientist Programme at ISSI, Bern executed by RT. Hospitality of ISSI  is thankfully recognized. We greatly appreciate the very critical and intriguing remarks of the two anonymous referees which contributed to substantial clarification and addition of the footnotes.

}

\end{document}